\documentclass[11pt,floatfix,preprint,showpacs]{revtex4}
\usepackage{epsfig}
\usepackage{graphicx}
\usepackage{dcolumn}
\usepackage{bm}
\usepackage{amsmath}
\usepackage{amssymb}
\usepackage{amsfonts}

\begin{document}

\title{Exact three-dimensional wave function and the on-shell t-matrix  for
the
sharply cut off Coulomb potential: failure of the standard renormalization
 factor}

\author{W.\ Gl\"ockle$^1$}

\author{J.~Golak$^2$}

\author{R.~Skibi\'nski$^2$}

\author{H.~Wita{\l}a$^2$}

\affiliation{$^1$Institut f\"ur theoretische Physik II,
Ruhr-Universit\"at Bochum, D-44780 Bochum, Germany}

\affiliation{$^2$M. Smoluchowski Institute of Physics, Jagiellonian
University, PL-30059 Krak\'ow, Poland}

\date{\today}

\begin{abstract}
The 3-dimensional wave function for a sharply cut-off Coulomb potential is
analytically derived.
The asymptotic form of the related scattering amplitude reveals a failure of
the standard renormalization
 factor which is believed to be generally valid for any type of screening.

\end{abstract}


\maketitle \setcounter{page}{1}

\section{Introduction}
\label{intro}

The long range behavior of the Coulomb force causes technical problems in
the scattering for more than
two particles. For instance the 3-body Faddeev kernel develops
singularities, which deny a direct
numerical approach. A way out has been searched in the past by starting with
a screened Coulomb
potential, which for instance in the context of the 3-body problem leads to
a screened 2-body  Coulomb t-matrix. In the limit of an infinite screening
radius it is claimed in the
literature~\cite{taylor1,taylor2,Alt78}
 that the on-shell
2-body t-matrix approaches the physical one except for an infinitely
oscillating phase factor, known
analytically. Thus removing that factor, called renormalization, the
physical result can be obtained.

As a basis for that approach work by Gorshkov~\cite{gor1,gor2},
 Ford~\cite{ford1964,ford1966} and Taylor~\cite{taylor1,taylor2} is
most often quoted. Gorshkov~\cite{gor1,gor2} regards
potential scattering on a Yukawa potential in the limit of the screening
radius going to infinity. He works
directly in 3 dimensions avoiding a partial wave decomposition. He sums up
the perturbation series to
infinite order. As a result he finds the limit for the wave function of a
Yukawa potential for an
infinite screening radius. That limit function equals the standard Coulomb
wave function multiplied by an
infinitely oscillating phase factor. Contrary to what is quoted in
Chen~\cite{chen72} he
has not achieved the
wave function for a Yukawa potential at an arbitrary screening radius but
only
its limiting form.

 The work by Ford~\cite{ford1964,ford1966} relies on a partial wave
decomposition. This leads to a
very difficult technical task
to handle the situation, when the orbital angular momentum  $l$ is about
$pR$,
where $p$ is the asymptotic
 wave number and $R$ the screening radius. This task is left unsolved  and
the
infinite sum in $l$ is  carried out without controlling the $l$-dependence
of certain correction
terms depending  on $R$.
 In other words the correction terms for given $l$ are assumed to remain
valid
also for the 3-dimensional
 objects. This leaves at least doubts about the rigorousness of that
approach.  The same is true for
 the investigations of Taylor~\cite{taylor1,taylor2}, where again a partial
wave decomposition is
the basis and the
 infinite sum over $l$ is carried through without control of its validity
for
the correction terms.

  In such a situation we felt that a rigorous analytical approach for a
sharply cut off Coulomb
potential carried through directly in 3 dimensions is in order.  This paper
delivers an analytical
solution for an arbitrary cut-off radius. Further we also provide an exact
expression for the corresponding
scattering amplitude (equivalent to the on-shell t-matrix). The paper is
organized as follows. In
section \ref{wavefunction} the wave function is derived. In section
\ref{scatamplitude} the scattering
amplitude and its limit for
vanishing screening is given. These purely analytical results are
confirmed by numerical studies
presented in section \ref{numresults}. In the Appendix we regard the much
simpler case for
s-wave scattering. We
summarize in section \ref{summary}.

\section{The wave function for a sharply  cut-off Coulomb potential}
\label{wavefunction}

Let us regard two equally charged particles with mass m. Then the 2-body
Schr\"odinger equation reads
\begin{eqnarray}
 ( - \nabla^2 - p^2 + \frac{me^2}{r} ) \Psi^{(+)} ( \vec r) = 0  ~.
\label{eq1}
\end{eqnarray}

It is well known that in parabolic coordinates
\begin{eqnarray}
u = r-z\\
v = r+z\\
\phi = tan^{-1} \frac{y}{x}
\end{eqnarray}
the partial differential equation factorizes and yields the solution
\begin{eqnarray}
\Psi^{(+)}(\vec r) = const~ e^{ i \vec p \cdot \vec r} {_1F_1} ( -
i \eta, 1, i ( pr - \vec p \cdot \vec r))\label{eq5}
\end{eqnarray}
with Somerfeld parameter $ \eta = \frac{ m e^2 }{ 2 p}$.

Now we switch to a sharply screened Coulomb potential
\begin{eqnarray}
V(r) = \Theta(R-r) \frac{e^2}{r}\label{eq6}
\end{eqnarray}
 and rewrite (\ref{eq1}) into the form of the Lippmann-Schwinger equation
\begin{eqnarray}
\Psi^{(+)}_R(\vec r) = \frac{ 1} { {(2 \pi)}^{ 3/2}} e^{ i \vec p \cdot \vec
r}
 - \frac{ m  }{ 4 \pi} \int d^3 r' \frac{      e^{ i p | \vec r - \vec r~'|
}
}{| \vec r - \vec r~'| }
\Theta( R-r') \frac{e^2}{r'}\Psi^{(+)}_R(\vec r~') ~.
\label{eq7}
\end{eqnarray}
This defines uniquely the wave function $\Psi^{(+)}_R(\vec r)$ for
a given cut-off radius $R$. Acting on (\ref{eq7}) with $ ( -
\nabla^2 - p^2)$ and using the well known property of the free
Greens function in the integral kernel one obtains the
Schr\"odinger equation
\begin{eqnarray}
 (- \nabla^2 - p^2  )\Psi^{(+)}_R(\vec r) = - m \Theta( R-r)
\frac{e^2}{r}\Psi^{(+)}_R(\vec r) ~.
\label{eq8}
 \end{eqnarray}
Thus for $ r < R$ one has to have
\begin{eqnarray}
\Psi_R^{(+)}(\vec r) = A e^{ i \vec p \cdot \vec r} {_1F_1} ( - i \eta, 1, i
( pr - \vec p \cdot \vec r))
\label{eq9}
\end{eqnarray}
with some to be determined constant $A$.
 The idea is therefore,  to insert that knowledge into the Lippmann-Schwinger  equation (\ref{eq7})
 leading to
\begin{eqnarray}
 & &  \Psi_R^{(+)}(\vec r) = \frac{ 1} { {(2 \pi)}^{ 3/2}} e^{ i \vec p
\cdot \vec r}\cr
 & - &  \frac{ m  }{ 4 \pi} \int d^3 r' \frac{ e^{ i p | \vec r - \vec r~'|
}
}{| \vec r - \vec r~'| }
\Theta( R-r') \frac{e^2 }{r'} A e^{ i \vec p \cdot \vec r~'} {_1F_1} ( - i
\eta, 1,
 i ( pr' - \vec p \cdot \vec r~')) ~.
\label{eq10}
\end{eqnarray}
If we choose $ r<R $ then also the left hand side is known and one
obtains the following identity
\begin{eqnarray}
& &   A e^{ i \vec p \cdot \vec r} {_1F_1} ( - i \eta, 1, i ( pr - \vec p
\cdot \vec r)
)  =   \frac{ 1} { {(2 \pi)}^{ 3/2}} e^{ i \vec p \cdot \vec r}\cr
 &  -&  \frac{ m  }{ 4 \pi} \int d^3 r' \frac{  e^{ i p | \vec r - \vec r~'|
} }{| \vec r - \vec r~'| }
\Theta( R-r') \frac{e^2 }{r'} A e^{ i \vec p \cdot \vec r~'} {_1F_1} ( - i
\eta, 1,
i ( pr' - \vec p \cdot \vec r~')) ~.
\label{eq11}
\end{eqnarray}
This provides the factor $A$. If $ A$ is known one can determine
the scattering amplitude $ f_R$ defined for $ r \to \infty$ by
\begin{eqnarray}
\Psi_R^{(+)}(\vec r) &\to & \frac{ 1} { {(2 \pi)}^{ 3/2}} e^{ i \vec p \cdot
\vec r}\cr
 &  + &  \frac{e^{ipr}}{r} A ( - \frac{ m}{ 4 \pi} )\int d^3 r' e^{ - ip
\hat r \cdot \vec r~'} \Theta( R-r')
\frac{e^2 }{r'} e^{ i \vec p \cdot \vec r~'} {_1F_1} ( - i \eta, 1, i ( pr'
-
\vec p \cdot \vec r~'))\cr
 & \equiv &  \frac{ 1} { {(2 \pi)}^{ 3/2}} e^{ i \vec p \cdot \vec r} +
\frac{e^{ipr}}{r}
f_R ~.
\label{eq12}
\end{eqnarray}

It is not difficult using properties of the confluent hypergeometric
function to show that the
corresponding LS
equation, for instance for a s-wave, is identically fulfilled as it should.
Doing that one can
read off the corresponding analytical expression for $A$. That calculation
is
deferred to the Appendix~\ref{apen1}.

The 3-dimensional case  is much harder. Let us choose $ \hat p =
\hat z$ and work with the
parabolic coordinates. Then (\ref{eq11}) turns into
\begin{eqnarray}
& &   A e^{ i \frac{p}{2} ( v-u)} {_1F_1} ( - i \eta, 1, i pu)
  =  \frac{ 1} { {(2 \pi)}^{ 3/2}} e^{ i \frac{p}{2} ( v-u)}\cr
 &   + &  A \frac{ e^2}{2}\int_0^{ 2R} du' e^{ -i\frac{p}{2}u'} {_1F_1} ( -
i \eta, 1, i pu')
\int_0^{ 2R-u'} dv'e^{ i\frac{p}{2}  v'}(-\frac{ m  }{ 4 \pi})
 \int_0^{ 2\pi} d\phi'\frac{  e^{ i p | \vec r - \vec r'| } }{| \vec r -
\vec r'| } ~. 
\label{eq13}
\end{eqnarray}
Since we want to determine just one factor $A$ one value of $ u$ and $v$ is
sufficient and we choose the
simplest case $ u=v=0$. Then the $ \phi'$ integration is trivial and one
obtains
\begin{eqnarray}
& & A =  \frac{ 1} { {(2 \pi)}^{ 3/2}}
  -   A \frac{ e^2 m}{2}\int_0^{ 2R} du'  {_1F_1} ( - i \eta, 1, i pu')
\int_0^{ 2R-u'} dv'e^{ ip v'}\frac{1}{ u' + v'} ~,
\label{eq14}
\end{eqnarray}
where we used $ {_1F_1} ( - i \eta, 1,0)=1$. Substituting $ u' = 2 R x, v' =
2 R y$ and defining $ A \equiv \tilde A \frac{ 1} { {(2 \pi)}^{ 3/2}}$ one
obtains
\begin{eqnarray}
\tilde A = 1 -\tilde  A \eta T \int_0^1 dx {_1F_1} ( - i \eta, 1,
iTx)\int_0^{1-x} dy e^{ iTy}\frac{1}{ x+y}
 \label{eq15}
\end{eqnarray}
with $ T \equiv 2 p R$.

Introducing  $ z \equiv iT$ let us define
\begin{eqnarray}
\tilde F(z) = 1 + \frac{ \eta z}{ i}\int_0^1 dx {_1F_1} ( - i
\eta, 1, zx)\int_0^{1-x} dy e^{ zy}\frac{1}{ x+y} ~.
\label{eq16}
\end{eqnarray}
Substituting $ zx = \tau, zy = \tau'$ we get
\begin{eqnarray}
\tilde F(z) = 1 - i \eta\int_0^z d\tau {_1F_1} ( - i \eta, 1,
\tau)\int_0^{z - \tau} d\tau' e^{\tau'}\frac{1}{ \tau + \tau'}~.
\label{eq17}
\end{eqnarray}
Then it follows
\begin{eqnarray}
\frac{ d\tilde F(z)}{ dz} & =  & -  \frac{i \eta}{z} e^{ z} \int_0^z d\tau
{_1F_1} ( - i \eta, 1, \tau) e^{- \tau}\\\label{eq18}
\frac{ d^2\tilde F(z)}{ dz^2} & = & \frac{i \eta}{z^2} e^{ z}
\int_0^z d\tau {_1F_1} ( - i \eta, 1, \tau) e^{- \tau}  - 
\frac{i \eta}{z} e^{ z} \int_0^z d\tau {_1F_1} ( - i \eta, 1,
\tau) e^{- \tau}\cr &  - &  \frac{i \eta}{z} e^{ z} \int_0^z d\tau
{_1F_1} ( - i \eta, 1, \tau) e^{- z}\cr & = & - \frac{1}{z}\frac{
d\tilde F(z)}{ dz} + \frac{ d\tilde F(z)}{ dz}-  \frac{i
\eta}{z}{_1F_1} ( - i \eta, 1, z) ~.
\label{eq19}
\end{eqnarray}
Consequently
\begin{eqnarray}
z\frac{ d^2\tilde F(z)}{ dz^2} + ( 1 -z)\frac{ d\tilde F(z)}{ dz}
=-  i \eta {_1F_1} ( - i \eta, 1, z) ~.
\label{eq20}
\end{eqnarray}
We add $ i \eta \tilde F(z) $ on both sides
\begin{eqnarray}
z\frac{ d^2\tilde F(z)}{ dz^2} + ( 1 -z)\frac{ d\tilde F(z)}{ dz} + i \eta
\tilde F(z)
=  i \eta ( \tilde F(z) - {_1F_1} ( - i \eta, 1, z)) ~.
\label{eq21}
\end{eqnarray}
The left side put to zero is the defining differential equation for $
{_1F_1} ( - i \eta, 1, z)$. Thus (\ref{eq21}) is fulfilled for
\begin{eqnarray}
\tilde F(z) ={_1F_1} ( - i \eta, 1, z)\label{eq22}
\end{eqnarray}
which also fixes the normalisation.

A rather lengthy sequence of analytical steps (not given) using an integral
representation, recurrence relations and further properties of
the confluent hypergeometric function yields the same result {\footnote {The
notes for that are available
from the authors.}.

Thus we obtain based on (\ref{eq15})
\begin{eqnarray}
\tilde A = 1 - \tilde A ({_1F_1} ( - i \eta, 1, iT)-1) ~.
\label{eq23}
\end{eqnarray}
The cancellation of  $ \tilde A$ on the left against $ \tilde A$
on the right is a verification that the LS equation (\ref{eq14})
at $ r=0$ is fulfilled, as it should and we end up with the exact relation
\begin{eqnarray}
\tilde A & = & \frac{1} { {_1F_1} (- i \eta,1,iT)} ~.
\label{eq24}
\end{eqnarray}

This is valid for any $T = 2 p R $ and therefore
\begin{eqnarray}
\Psi_R^{(+)}(\vec r) =  \frac{1}{ ( 2 \pi)^{ 3/2}}\frac{1} {
{_1F_1} (- i \eta,1,iT)}   e^{ i \vec p \cdot \vec r} {_1F_1} ( -
i \eta, 1, i ( pr - \vec p \cdot \vec r)) \label{eq25}
\end{eqnarray}
is exactly fulfilled for $ r < R$, inside the range of the potential.
To the best of our knowledge this is the first time that this has
been achieved.

At the same time it provides due to (\ref{eq12}) the exact
expression for the scattering amplitude $ f_R$ or the on-shell
t-matrix element for a sharply cut-off Coulomb potential. This
will be dealt with in the next section.

\section{The scattering amplitude}
\label{scatamplitude}

The starting point  due to (\ref{eq12}) with $ \tilde f_R = \frac{1}{( 2
\pi)^ { 3/2}} f_R$ is
 \begin{eqnarray}
   \tilde f_R = \tilde A ( - \frac{ m}{ 4 \pi} )
\int^R d^3 r' e^{ - i p \hat r \cdot \vec r~'} \frac{ e^2}{ r'} e^{
i \vec p \cdot \vec r~'}  F( - i \eta, 1, ( pr' - \vec p \cdot \vec
r~')) ~.
\label{eq26}
\end{eqnarray}
We use the general  integral representation of $  F(\alpha,\beta,z)$
\begin{eqnarray}
 {_1F_1} ( \alpha,\beta,z) = C(\alpha,\beta)
\int_{ \Gamma} dt e^ { zt} t^ { \alpha-1} ( 1-t)^ { \beta- \alpha
-1}\label{eq27}
\end{eqnarray}
where the path $ \Gamma$ encircles the logarithmic cut between $
t=0$ and $ t=1$ in the positive sense and the prefactor is
\begin{eqnarray}
C( \alpha,\beta) = \frac{ \Gamma( \beta)}{ \Gamma( \alpha) \Gamma( \beta-
\alpha) }
\frac{ 1}{ 1 - e^ { 2 \pi i( \beta - \alpha)}} ~.
\label{eq28}
\end{eqnarray}

Inserting (\ref{eq27}) into (\ref{eq26}) yields
\begin{eqnarray}
\tilde f_R & = &  \tilde A ( - \frac{ m e^2}{ 4 \pi} )C( - i
\eta,1) \int_{ \Gamma} dt ( \frac{1-t}{t})^{ i \eta} \frac{ 1}{ t}
  \int^R d^3 r' e^{ - i p \hat r \cdot \vec r~'}\frac{1}{ r'} e^{ i
    \vec p
\cdot \vec r'}
e^{ i( pr' - \vec p \cdot \vec r~')t} ~.
\label{eq29}
\end{eqnarray}
The $ \vec r~'$ integral is straightforward and one obtains
\begin{eqnarray}
   \tilde f_R & = &  \tilde A ( - \frac{ m e^2}{ 2 p^2 \alpha} )C( - i
\eta,1)
\int_{ \Gamma} dt ( \frac{1-t}{t})^{ i \eta} \frac{1}{ t (
1-t)}\cr & & (1 +e^{ i \tilde R t} ( it \frac{ sin  \tilde R
\sqrt{ t^2 + 2 ( 1-t) \alpha}}{ \sqrt{ t^2 + 2 ( 1-t) \alpha}} -
cos \tilde R \sqrt{ t^2 + 2 ( 1-t) \alpha}))\label{eq30}
\end{eqnarray}
where $ \alpha$ contains the dependence on the scattering angle $ \theta$
\begin{eqnarray}
\alpha = 1 - \hat p \cdot \hat r = 2 sin^ 2 \frac{ \theta}{2} ~,
\label{eq31}
\end{eqnarray}
$ C( - i \eta,1) = \frac{ -i}{ 2 \pi} e^ { \pi \eta} $, 
and $ \tilde R \equiv p R$.
The $"1"$ in the bracket does not contribute since
\begin{eqnarray}
 \int_{ \Gamma} dt ( \frac{1-t}{t})^{ i \eta} \frac{1}{ t ( 1-t)}
=0 ~.
\label{eq32}
\end{eqnarray}
Thus we obtain the intermediate result
\begin{eqnarray}
\tilde f_R & = &  - \tilde A  \frac{ \eta}{ \alpha p} C( - i
\eta,1)  [ i \int_{ \Gamma} dt ( \frac{1-t}{t})^{ i \eta}
\frac{1}{ 1-t} e^{ i \tilde R t} \frac{sin  \tilde R \sqrt{ t^2 +
2 ( 1-t) \alpha}}{ \sqrt{ t^2 + 2 ( 1-t) \alpha}} \cr  
& - & \int_{
\Gamma} dt (\frac{1-t}{t})^{ i \eta} \frac{1}{ t( 1-t)} e^{ i
\tilde R t}  cos \tilde R \sqrt{ t^2 + 2 ( 1-t) \alpha}] ~.
\label{eq33}
\end{eqnarray}

In the following we choose the path of integration $ \Gamma$ as
depicted in Fig.~\ref{fig1} with small circles around $ t=1$ and $ t=0$ of
vanishingly small radius $ \epsilon$ and two straight integration
lines between $ t= \epsilon$ and $ t = 1-\epsilon$ above and below
the logarithmic cut. The phases are : $arg (t) =0$ and $arg( 1-t)
= \pi$ for $t= 1 + \epsilon$. The rest follows by continuity:
$arg( 1-t)= 2 \pi$ along the upper rim of the cut, $arg(t) = 2 \pi$ along
the
lower rim and $arg( 1-t) = 3 \pi$ back again at $t= 1+ \epsilon$.
The phase of $\frac{1-t}{t}$ does not change after a full sweep of
$\Gamma$, of course.

In this manner the integrals in (\ref{eq33}) can be split into 4 pieces.
Let us define
  \begin{eqnarray}
B &\equiv  &  \int_{\Gamma} dt ( \frac{1-t}{t})^{ i \eta} \frac{1}{ 1-t} e^{
i
\tilde R t} ( i \frac{sin  \tilde R \sqrt{ t^2 + 2 ( 1-t)
\alpha}}{ \sqrt{ t^2 + 2 ( 1-t) \alpha}}
 - \frac{cos \tilde R \sqrt{ t^2 + 2 ( 1-t) \alpha}}{t}) ~.
\label{eq34}
\end{eqnarray}
Then
\begin{eqnarray}
B = \int_{zero} + \int_{ \epsilon}^ { 1 - \epsilon} + \int_{one} +
\int_{ 1-\epsilon}^ { \epsilon} ~.
\label{eq35}
\end{eqnarray}
It simply follows
\begin{eqnarray}
& & \int_{ 1-\epsilon}^ { \epsilon}dt  + \int_{\epsilon}^ {1-
\epsilon} dt =  ( 1 - e^ { - 2 \pi \eta}) \int_{\epsilon}^
{1- \epsilon} dt ( \frac{  1-t }{ t})^ { i \eta}
 \frac{ 1}{1-t}   e^{ i \tilde R t}\cr
& & ( i \frac{sin  \tilde R \sqrt{ t^2 + 2 ( 1-t) \alpha}}{ \sqrt{
t^2 + 2 ( 1-t) \alpha}}
 - \frac{cos \tilde R \sqrt{ t^2 + 2 ( 1-t) \alpha}}{t}) ~.
\label{eq36}
\end{eqnarray}

In order to remove the pole singularities at $ t=0$ and $ t=1$ we
split the integration interval into two parts
 \begin{eqnarray}
\int_{\epsilon}^ {1- \epsilon} dt= \int_{\epsilon}^ {1/2} dt+
\int_{1/2}^ {1-\epsilon} dt ~.
\label{eq37}
\end{eqnarray}
Of course the value $ 1/2$ could be replaced by any number $a$
between $ t= \epsilon$ and $ t=1- \epsilon$ without changing the
result.

Thus
\begin{eqnarray}
\int_{\epsilon}^ {1- \epsilon} dt
 & = &
 i \int_{\epsilon}^ {1/2} dt (1-t)^ { i \eta -1}
 t^ { - i \eta} e^{ i \tilde R t} \frac{sin  \tilde R \sqrt{ t^2 + 2 ( 1-t)
\alpha}}
{ \sqrt{ t^2 + 2 ( 1-t) \alpha}}\cr
&   + &  i  \int_{1/2}^ {1 -
\epsilon} dt(1-t)^ { i \eta -1}
 t^ { - i \eta} e^{ i \tilde R t} \frac{sin  \tilde R \sqrt{ t^2 + 2 ( 1-t)
\alpha}}
{ \sqrt{ t^2 + 2 ( 1-t) \alpha}}\cr & -& \int_{\epsilon}^ {1/2} dt
t^ { - i \eta -1} ( 1-t)^ { i \eta -1} e^{ i \tilde R t}
 cos \tilde R \sqrt{ t^2 + 2 ( 1-t) \alpha}\cr
& - & \int_{1/2}^ {1- \epsilon} dt ( 1-t)^ { i \eta -1} t^ { - i
\eta -1}  e^{ i \tilde R t}
 cos \tilde R \sqrt{ t^2 + 2 ( 1-t) \alpha} ~.
\label{eq38}
 \end{eqnarray}

Now we perform partial integrations such that $ \epsilon
 \rightarrow 0$ can be taken:
\begin{eqnarray}
 \int_{\epsilon}^ {1- \epsilon} dt
 & = &  i \int_0^ {1/2} dt (1-t)^ { i \eta -1} t^ { - i \eta} e^{ i \tilde R
t}
 \frac{sin  \tilde R \sqrt{ t^2 + 2 ( 1-t) \alpha}}{ \sqrt{ t^2 + 2 ( 1-t)
\alpha}}\cr
& + &  i[ \frac{-1}{ i \eta} ( 1 - t)^ { i \eta} t^ { - i \eta}
e^{ i \tilde R t}
 \frac{sin  \tilde R \sqrt{ t^2 + 2 ( 1-t) \alpha}}{ \sqrt{ t^2 + 2 ( 1-t)
\alpha}}|_ {1/2}^ { 1-
\epsilon}\cr
 & + &  \frac{1}{i\eta} \int_{1/2}^ {1 - \epsilon} dt(
1 - t)^ { i \eta} \frac{d}{dt} ( t^ { - i \eta}
 e^{ i\tilde R t}
 \frac{sin  \tilde R \sqrt{ t^2 + 2 ( 1-t) \alpha}}{ \sqrt{ t^2 + 2 ( 1-t)
 \alpha}}]\cr
  &-& [ \frac{1}{ - i \eta} t^ { - i \eta} (1-t)^ { i \eta -1} e^{ i \tilde
R t}
cos \tilde R \sqrt{ t^2 + 2 ( 1-t) \alpha}|_{\epsilon}^ {1/2}\cr
 &
+ &  \frac{1}{ i \eta} \int_0^ {1/2} dt t^ { - i \eta}
\frac{d}{dt}( (1-t)^ { i \eta -1}
 e^{ i \tilde R t} cos \tilde R \sqrt{ t^2 + 2 ( 1-t) \alpha})]\cr
&  -&  [ \frac{1}{ - i \eta}(1-t)^ { i \eta} t^ { - i \eta-1} e^{
i \tilde R t} cos \tilde R \sqrt{ t^2 + 2 ( 1-t) \alpha}|_{1/2}^ {
1 - \epsilon}\cr
 &  + &  \frac{1}{ i \eta}\int_{1/2}^ 1 dt (1-t)^
{ i \eta}\frac{d}{dt}(t^ { - i \eta-1} e^{ i \tilde R t} cos
\tilde R \sqrt{ t^2 + 2 ( 1-t) \alpha})] ~.
\label{eq39}
\end{eqnarray}

After some lengthy algebra one obtains
\begin{eqnarray}
 \int_{\epsilon}^ { 1 - \epsilon} dt &=&    \frac{1}{i  \eta}
\epsilon^ { i \eta}  -   \frac{ 1}{ i \eta} \epsilon^ { - i \eta} cos \tilde
R \sqrt{ 2 \alpha} 
   +   \frac{1}{  \eta} e^ { i \frac{\tilde R}{2}} \frac{sin  \tilde R
\sqrt{ 1/4 +  \alpha}}
 { \sqrt{ 1/4 +\alpha}}\cr
 & + & i \int_0^ {1/2} dt (1-t)^ { i \eta -1} t^ { - i \eta} e^{ i \tilde R
t} \frac{sin  \tilde R \gamma}{ \gamma} 
 +   \frac{i \eta -1}{  i\eta} \int_0^ { 1/2} dt t^ { - i \eta} ( 1-t)^ {
i \eta -2}
 e^ { i \tilde R t} cos \tilde R \gamma\cr
& - & \frac{ 1}{ \eta} \int_{1/2}^ 1 dt( 1 - t)^ { i \eta}t^ { - i \eta}
e^{ i\tilde R t}\frac{sin  \tilde R \gamma} { \gamma^ 3} ( t - \alpha)
  -  i \int_{1/2}^ 1 dt( 1 - t)^ { i \eta}  t^ { - i \eta -1}  e^{ i\tilde
R t}
 \frac{sin  \tilde R \gamma} { \gamma}\cr
 & + &  \frac{ i \eta +1}{ i \eta}  \int_{1/2}^ 1 dt ( 1-t)^ { i \eta} t^ {
- i \eta -2}
 e^ { i \tilde R t} cos \tilde R \gamma\cr
& + & \frac{ \tilde R}{\eta} [- \int_0^ { 1/2} dt t^ { - i \eta} ( 1-t)^ { i
\eta -1}
 e^ { i \tilde R t} cos \tilde R \gamma
 +   \frac{ 1}{i}  \int_0^ { 1/2} dt t^ { - i \eta} ( 1-t)^ { i \eta -1}
e^ { i \tilde R t}
  sin  \tilde R \gamma  \frac{ t - \alpha}{ \gamma}\cr
&  + &  i  \int_{1/2}^ 1 dt( 1 - t)^ { i \eta } t^ { - i \eta }
 e^{ i\tilde R t} \frac{sin  \tilde R \gamma} { \gamma} 
 +   \int_{1/2}^ 1 dt( 1 - t)^ { i \eta }   t^ { - i \eta} e^ { i \tilde R
t}
 \frac{ cos \tilde R \gamma} { \gamma}  \frac{ t - \alpha}{ \gamma}\cr
& - & \int_{1/2}^ 1 dt ( 1-t)^ { i \eta} t^ { - i \eta -1}e^ { i \tilde R t}
cos \tilde R \gamma 
 +  \frac{1}{i} \int_{1/2}^ 1 dt ( 1-t)^ { i \eta} t^ { - i \eta -1} e^ {
i \tilde R t}
 sin  \tilde R \gamma \frac{ t - \alpha}{ \gamma}]\label{eq40}
\end{eqnarray}
with $ \gamma = \sqrt{ t^2 + 2 ( 1-t) \alpha}$.

It is straightforward to evaluate the two integrals around $ t=0$
and $ t=1$:
\begin{eqnarray}
& & \int_{zero} dt ( \frac{1-t}{t})^{ i \eta} \frac{1}{ 1-t} e^{ i \tilde R
t}
  ( i \frac{sin  \tilde R \sqrt{ t^2 + 2 ( 1-t) \alpha}}{ \sqrt{ t^2 + 2
( 1-t) \alpha}}
 - \frac{cos \tilde R \sqrt{ t^2 + 2 ( 1-t) \alpha}}{t})\cr
 & = & - i cos \tilde R \sqrt{ 2 \alpha} \epsilon^ { - i \eta} \frac{1}{
\eta}( 1 - e^ { - 2 \pi
\eta}) ~,
\label{eq41}
\end{eqnarray}
\begin{eqnarray}
& & \int_{one} dt ( \frac{1-t}{t})^{ i \eta} \frac{1}{ 1-t} e^{ i
    \tilde R t}
  ( i \frac{sin  \tilde R \sqrt{ t^2 + 2 ( 1-t)
\alpha}}{ \sqrt{ t^2 + 2 ( 1-t) \alpha}} - \frac{cos \tilde R \sqrt{ t^2 + 2
( 1-t) \alpha}}{t})\cr
& = &  \frac{ i}{ \eta} \epsilon^ { i \eta}( 1 - e^ { - 2 \pi
\eta}) ~.
\label{eq42}
\end{eqnarray}
The $\epsilon$-dependent terms cancel in (\ref{eq40}) (multiplied by $ (
1 - e^ { - 2 \pi \eta})$ ), 
 (\ref{eq41}) and (\ref{eq42}) as they
should and one obtains the finite result
\begin{eqnarray}
B & = & ( 1 - e^ { - 2 \pi \eta})
  [    \frac{1}{  \eta} e^ { i \frac{\tilde R}{2}} \frac{sin  \tilde R
\sqrt{ 1/4 +  \alpha}}
{ \sqrt{ 1/4 +\alpha}} +  i \int_0^ {1/2} dt (1-t)^ { i \eta
-1} t^ { - i \eta} e^{ i \tilde R t}
 \frac{sin  \tilde R \gamma}{ \gamma}\cr
& + &  \frac{i \eta -1}{  i\eta} \int_0^ { 1/2} dt t^ { - i \eta}
( 1-t)^ { i \eta -2}
 e^ { i \tilde R t} cos \tilde R \gamma
 -  \frac{ 1}{ \eta} \int_{1/2}^ 1 dt( 1 - t)^ { i \eta}t^ { - i
\eta}e^{ i\tilde R t}\frac{sin  \tilde R \gamma} { \gamma^ 3} ( t
- \alpha)\cr & - & i \int_{1/2}^ 1 dt( 1 - t)^ { i \eta}  t^ { - i
\eta -1}  e^{ i\tilde R t} \frac{sin  \tilde R \gamma} {
\gamma} +   \frac{ i \eta +1}{ i \eta}  \int_{1/2}^ 1 dt (
1-t)^ { i \eta} t^ { - i \eta -2}
 e^ { i \tilde R t} cos \tilde R \gamma\cr
& + & \frac{ \tilde R}{\eta} [- \int_0^ { 1/2} dt t^ { - i \eta} (
1-t)^ { i \eta -1}  e^ { i \tilde R t} cos \tilde R \gamma
+   \frac{ 1}{i}  \int_0^ { 1/2} dt t^ { - i \eta} ( 1-t)^ { i \eta -1}
e^ { i \tilde R t}
   sin  \tilde R \gamma  \frac{ t - \alpha}{ \gamma}\cr
&  + &  i  \int_{1/2}^ 1 dt( 1 - t)^ { i \eta } t^ { - i \eta } e^{ i\tilde
R t} \frac{sin  \tilde R \gamma} { \gamma} 
+   \int_{1/2}^ 1 dt( 1 - t)^ { i \eta }   t^ { - i \eta}
 e^ { i \tilde R t} \frac{ cos \tilde R \gamma} { \gamma}  \frac{ t -
\alpha}{ \gamma}\cr
& - & \int_{1/2}^ 1 dt ( 1-t)^ { i \eta} t^ { - i \eta -1}e^ { i \tilde R t}
cos \tilde R \gamma 
 +  \frac{1}{i} \int_{1/2}^ 1 dt ( 1-t)^ { i \eta} t^ { - i \eta -1} e^ {
i \tilde R t}
 sin  \tilde R \gamma \frac{ t - \alpha}{ \gamma}]] ~.
\label{eq43}
\end{eqnarray}
This together with (\ref{eq33})-(\ref{eq36}) is an exact expression for the
scattering amplitude for an arbitrary cut-off radius $R$.

But of course we are interested only in its asymptotic limit $ R \to
\infty$.

It is advisable to introduce $e_{\pm} = e^ { i \tilde R ( t \pm
  \gamma)}$ and to rearrange (\ref{eq43}). We regard first the pieces
  explicitly proportional to $ \tilde R$ in (\ref{eq43})
\begin{eqnarray}
   & &  \frac{ \tilde R}{2\eta} [- \int_0^ { 1/2} dt ( \frac{1-t}{t})^ { i
\eta} \frac{1}{1-t}
( e_+ + e_-) 
 -     \int_0^ { 1/2} dt ( \frac{1-t}{t})^ { i
\eta}\frac{1}{1-t}  ( e_+ - e_-) \frac{ t - \alpha}{ \gamma} \cr 
&+&    \int_{1/2}^ 1 dt ( \frac{1-t}{t})^ { i \eta}  \frac{ e_+ -
e_-} { \gamma}  +   \int_{1/2}^ 1 dt ( \frac{1-t}{t})^ { i
\eta} \frac{e_+ + e_-} { \gamma}  \frac{ t - \alpha}{ \gamma}\cr 
&-& \int_{1/2}^ 1 dt ( \frac{1-t}{t})^ { i \eta}\frac{1}{t} (e_+ +
e_-)  -   \int_{1/2}^ 1 dt ( \frac{1-t}{t})^ { i
\eta}\frac{1}{t} ( e_+ - e_-) \frac{ t - \alpha}{ \gamma}] ~. 
\label{eq44}
\end{eqnarray}
Leading terms will arise from the boundaries of integration $ t=0$,
 $t=1/2$, and $  t=1$, where the $ t=1/2 $ contributions have to cancel in
the total expression. We use the standard method of steepest
descent~\cite{steepest}  and expand around boundaries of integration. For
example at $ t=0$
\begin{eqnarray}
e_{\pm} &  = &    e^ { \pm i \tilde R \sqrt{ 2 \alpha}} e^ { i
\tilde R  t( 1 \mp \sqrt{ \frac{ \alpha}{2}})} ( 1 + O( t))\label{eq45}
\end{eqnarray}
and corresponding expressions for the remaining parts of the
integrand. One obtains
\begin{eqnarray}
& & \frac{ \tilde R}{2\eta}  [- \int_0 dt ( \frac{1-t}{t})^ { i \eta}
\frac{1}{1-t} ( e_+ + e_-) 
  -  \int_0 dt ( \frac{1-t}{t})^ { i \eta}\frac{1}{1-t}  ( e_+ - e_-)
\frac{ t - \alpha}{ \gamma}] \cr
 & \to & \frac{ 1}{2 \eta} ( - i e^ { \frac{ \pi}{2} \eta}
 \Gamma( 1 - i \eta) \tilde R^ { i \eta} 
   ( e^ { 2 i \tilde R  sin \frac{ \theta}{2}} (1 - sin \frac{
\theta}{2})^ { i \eta}
 + e^ { - 2 i \tilde R  sin \frac{ \theta}{2}} (1 +  sin \frac{ \theta}{2})^
{ i \eta}) ~.
\label{eq46}
\end{eqnarray}
Correspondingly we proceed at the upper limit of integration $ t=1$ and  it
turns
out that the $ e_+$ part decreases as $ O( \frac{1}{\tilde R})$
and only the $ e_-$ part survives as
\begin{eqnarray}
& & \frac{ \tilde R}{2\eta}  [  \int^ 1 dt ( \frac{1-t}{t})^ { i \eta}
\frac{ e_+ - e_-}  { \gamma} 
  +  \int^ 1 dt ( \frac{1-t}{t})^ { i \eta} \frac{e_+ + e_-} { \gamma}
\frac{ t - \alpha}{ \gamma} \cr & - & \int^ 1 dt ( \frac{1-t}{t})^ { i
\eta}\frac{1}{t} (e_+ + e_-)
  -  \int^ 1 dt ( \frac{1-t}{t})^ { i \eta}\frac{1}{t} ( e_+ - e_-) \frac{
t - \alpha}{ \gamma}] \cr 
& \to & \frac{i }{ \eta} ( 2 \tilde R)^ { - i \eta} (sin^ 2 \frac{
\theta}{2})^ { -i \eta}
 e^ { \frac{ \pi}{2} \eta} \Gamma( 1 + i \eta) ~. 
\label{eq47}
\end{eqnarray}

The remaining pieces resulting from the integration limits $t=1/2$ yield
\begin{eqnarray}
& & \frac{ \tilde R}{2\eta}[- \int^ { 1/2} dt t^ { - i \eta} ( 1-t)^ { i
\eta -1} ( e_+ + e_-) 
  -     \int^ { 1/2} dt t^ { - i \eta} ( 1-t)^ { i \eta -1}( e_+ - e_-)
\frac{ t - \alpha}{ \gamma}\cr
&  + &    \int_{1/2} dt( 1 - t)^ { i \eta } t^ { - i \eta } \frac{ e_+ -
e_-} { \gamma} 
 +   \int_{1/2} dt( 1 - t)^ { i \eta }   t^ { - i \eta}\frac{e_+ + e_-} {
\gamma}
  \frac{ t - \alpha}{ \gamma}\cr
& - & \int_{1/2} dt ( 1-t)^ { i \eta} t^ { - i \eta -1}e^ { i \tilde R t}
(e_+ + e_-)
 -   \int_{1/2} dt ( 1-t)^ { i \eta} t^ { - i \eta -1} ( e_+ - e_-) \frac{
t - \alpha}{ \gamma}]\cr
 & \to &  - \frac{1}{\eta} \frac{1}{ \sqrt{ 1/4 + \alpha}} e^ { i
\frac{\tilde R}{2} }
 sin \tilde R \sqrt{ 1/4 + \alpha} ~. 
\label{eq48}
\end{eqnarray}
This cancels exactly against the first term in (\ref{eq43}) after
multiplication by $ ( 1 - e^ { - 2 \pi
\eta})$, as it should.

The terms in (\ref{eq43}) not directly proportional to $ \tilde R$ decrease
like $ O( \frac{1}{ \tilde R})$. Finally the contributions from
the interior of the integration intervals decay faster as can be
seen by deforming the path of integration into the upper half
plane, where $ e^{\pm}$ is exponentially damped.

Thus we are left with the leading asymptotic expression
\begin{eqnarray}
 B &\to &    ( 1 - e^ { - 2 \pi \eta})\frac{ i}{ 2 \eta} e^ { \frac{
\pi}{2} \eta}
 (~ - \Gamma( 1 - i \eta) \tilde R^ { i \eta} (~ e^ { 2 i \tilde R sin
\frac{ \theta}{2}} (1 - sin \frac{ \theta}{2})^ { i \eta}
 + e^ { - 2 i \tilde R  sin \frac{ \theta}{2}} (1 +  sin \frac{ \theta}{2})^
{ i \eta}~) \cr
& + & 2  ( 2 \tilde R)^ { - i \eta} (sin^ 2 \frac{ \theta}{2})^ { -i \eta}
\Gamma( 1 + i
\eta)~) ~.
\label{eq49}
\end{eqnarray}
This is now to be combined with (\ref{eq33}). Using (\ref{eq28}),
\begin{eqnarray}
\frac{\Gamma( 1 + i \eta)}{\Gamma( 1 - i \eta)} \equiv e^ { 2 i
\sigma_0}\label{eq50}
\end{eqnarray}
and the asymptotic form of $ \tilde A$
\begin{eqnarray}
\tilde A \to e^ {-\frac{\pi}{2} \eta} \, T^ {- i \eta} \, { \Gamma(1 + i \eta)}
\label{eq511}
\end{eqnarray}
based on the asymptotic form \cite{abromowitz}
\begin{eqnarray}
{_1F_1} (\alpha,\beta,z) \to \frac{e^ {\pm i \pi \alpha} z^ { -
\alpha}}{ \Gamma( \beta - \alpha)}
+ \frac{ e^ z z^ { \alpha - \beta}}{ \Gamma( \alpha)} + O( \frac{1}{ |
z|}) ~, 
\label{abrom}
\end{eqnarray}
 we get
\begin{eqnarray}
  \tilde f_R &  = & - ( 2 \tilde R)^ { -2 i \eta} e^ { 2 i \sigma_0}
\frac{\eta}{ 2  p} \frac{  (sin^ 2 \frac{ \theta}{2})^ { -i
\eta}}{ sin^ 2 \frac{ \theta}{2}}  \cr
 & + &   \frac{\eta}{2 \alpha p}  ( e^ { 2 i \tilde R  sin \frac{
\theta}{2}}
(\frac{(1 - sin \frac{ \theta}{2})}{2})^ { i \eta}
 + e^ { - 2 i \tilde R sin \frac{ \theta}{2}} (\frac{(1 +  sin \frac{
\theta}{2})}{2})^ { i
\eta}) ~. 
\label{eq51}
\end{eqnarray}
Now the physical Coulomb scattering amplitude is
\begin{eqnarray}
A_c( \theta) = - \frac{\eta}{ 2  p}\frac{  (sin^ 2 \frac{ \theta}{2})^ { -i
\eta}}{ sin^ 2 \frac{ \theta}{2}}
 e^ { 2 i \sigma_0}\label{eq52}
\end{eqnarray}
and we end up with
\begin{eqnarray}
\tilde f_R &  = &  ( 2 \tilde R)^ { -2 i \eta} A_c( \theta) 
 +    \frac{\eta}{4 p  sin^ 2 \frac{ \theta}{2}}  ( e^ { 2 i \tilde R  sin
\frac{ \theta}{2}}
 (\frac{(1 - sin \frac{ \theta}{2})}{2})^ { i \eta}
 + e^ { - 2 i \tilde R sin \frac{ \theta}{2}} (\frac{(1 +  sin \frac{
\theta}{2})}{2})^ { i
\eta})\cr
&=& [e^{-2i\eta ln(2\tilde R)} -\frac {1} {2} e^{i\eta ln sin^2\frac
   {\theta} {2} -2i{\sigma}_0} (e^{2i\tilde R sin \frac
   {\theta} {2} +i\eta ln \frac {1-sin \frac
   {\theta} {2}} {2}} + e^{-2i\tilde R sin \frac
   {\theta} {2} +i\eta ln \frac {1+sin \frac   {\theta} {2}} {2}})]
A_c( \theta) 
 ~.
\label{eq53}
\end{eqnarray}

The first term is the result expected from the literature
\cite{taylor1,Alt78}  and references therein. As
\cite{kamada05} has shown,
 the diverging phase factor $ e^ { - 2i \Phi_R(p)}$ in case of an often used
form of screening the Coulomb potential
\begin{eqnarray}
V_R(r) = \frac{e^ 2}{r} e^ { - (\frac{r}{R})^ n}\label{eq54}
\end{eqnarray}
using the prescription of \cite{taylor1} turns out to be
\begin{eqnarray}
\Phi_R(p) = \eta [ ln (2 p R) - C/n ] 
\label{eq55}
\end{eqnarray}
with the Euler number $C$. For $ n \to \infty$ one recovers the sharp
cut-off, which we consider in
this paper.  This expectation for the screening limit agrees with the first
term in (\ref{eq53}) but not
with the necessity of adding a second term. Therefore the derivations in the
literature based on partial
wave decomposition must be  incomplete. Whether this is also true for a
finite value $n$ in (\ref{eq54})
 remains to be seen.

\section{Numerical results}
\label{numresults}

We performed a number of numerical tests to check the basic points in
 the derivation of the sharp cut off Coulomb wave function (\ref{eq25}) and
 the asymptotic scattering amplitude (\ref{eq53}). 

First we checked numerically how well the solution (\ref{eq24})
fulfills equation (\ref{eq15}). In Table~\ref{table1} 
 the left and right sides of
(\ref{eq15}) are shown for a number of cut-off radii $R$ for pp scattering
with $E_p^{lab}=13$~MeV. The right side was obtained by a direct
two-dimensional numerical integration over $x$ and $y$. The very good
agreement up to four significant digits is seen. 

We also compared at the same energy the exact expression for $\tilde A$
as given in (\ref{eq24}) with its asymptotic form (\ref{eq511}) at a
number of screening radii. The results are shown in Fig.~\ref{fig2}
and Table~\ref{table2}. The oscillating behavior seen in real and
imaginary parts of exact $\tilde A$ (solid lines in Fig.~\ref{fig2})
 gradually diminishes with increasing cut-off radius $R$. These
 oscillations are absent in the asymptotic form for $\tilde A$ (dashed
 lines in Fig.~\ref{fig2}). The asymptotic form for $\tilde A$
 approaches its exact value at $R \approx 50$~fm as can be seen in
 Fig.~\ref{fig2} and in the third column of Table~\ref{table2} where
 the ratio of $\tilde A / \tilde A_{approx}$ is given.

To check the quality of our renormalization factor (\ref{eq53}) we applied
it directly to the numerical solutions of the Lippmann-Schwinger
equation  for the sharp cut off Coulomb potential with different
cut-off radii. 

In the case of a short-ranged potential $V$ two-body scattering
is described by the solution of the Lippmann-Schwinger equation
\begin{eqnarray}
 T(z) = V + V \frac1{z - H_0} T(z),
\label{genLSE}
\end{eqnarray}
where $V$ is the two-body potential, $H_0$ is the free Hamiltonian and
$T(z)$ the transition operator. In momentum space Eq.~(\ref{genLSE}) takes
the form of an integral equation for the matrix elements of the transition
operator $ \langle {\vec q}^{\ \prime} \mid T(z) \mid  {\vec q} \rangle
\equiv T(  {\vec q}^{\ \prime} ,  {\vec q} \, )$. In this equation matrix
elements of the potential $V$ are used $ \langle {\vec q}^{\ \prime} \mid V
\mid  {\vec q} \rangle
\equiv V(  {\vec q}^{\ \prime} , {\vec q} \, )$. In our case both $  V(
{\vec q}^{\ \prime} , {\vec q} \, )$ and
 $T(  {\vec q}^{\ \prime} ,  {\vec q} \, )$
depend only on the magnitudes $ q^\prime \equiv \mid  {\vec q}^{\ \prime}
\mid $,
$ q \equiv \mid  {\vec q} \mid $ and the cosine of the angle between $
{\vec q} $ and $ {\vec q}^{\ \prime} $, $  {\hat q}^{\ \prime}
\cdot {\hat q}$:
\begin{eqnarray}
    V(  {\vec q}^{\ \prime} , {\vec q} \, ) =
      V( q^\prime , q,  {\hat q}^{\ \prime} \cdot {\hat q} )
\end{eqnarray}
\begin{eqnarray}
    T(  {\vec q}^{\ \prime} , {\vec q} \, ) =
      T( q^\prime , q,  {\hat q}^{\ \prime} \cdot {\hat q} ) .
\end{eqnarray}
(Note we dropped the dependence on the parameter $z$.)
As a consequence the Lippmann-Schwinger equation can be written as a
two-dimensional integral equation \cite{elster1998}
\begin{eqnarray}
 T( q^\prime , q, x^\prime) = \frac1{2 \pi} v( q^\prime , q, x^\prime, 1 )
+\int\limits_0^\infty d q''  {q''}^{\, 2}
 \int\limits_{-1}^1 d x'' v( q^\prime , q'' , x^\prime, x'' )  \frac1{ z -
\frac{{q''}^{\, 2}}{m}}
 T( q'' , q, x'' ) ,
\label{2dimLSE}
\end{eqnarray}
where \begin{eqnarray}
v( q^\prime , q , x^\prime, x )  =
 \int\limits_{0}^{2 \pi} d \varphi V( q^\prime, q, x^\prime x + \sqrt{ 1 -
{x^\prime}^{\, 2} } \sqrt{ 1 - {x}^{\, 2} } \cos \varphi ) \label{smallv}
\end{eqnarray}
and $m$ is the reduced mass of the system.

For the sharply screened Coulomb potential of the range $R$ considered in
this paper \begin{eqnarray}
V ( q^\prime, q , x ) = \frac{e^2}{ 2 \pi^2 } \,
               \frac{ 1 - \cos ( Q R ) }{ Q^2 },
\end{eqnarray}
where $ Q \equiv \sqrt{ {q^\prime}^{\, 2} + q^2 - 2 q^\prime \, q x }$.
However, the integral over $\varphi$ in Eq.~(\ref{smallv}) cannot be carried
out analytically.

It is clear that $ V ( q^\prime, q , x ) $ shows a highly oscillatory
behavior, especially for large $R$. Thus solving the two-dimensional
equation (\ref{2dimLSE}) is a difficult numerical problem. We were
interested in solutions for positive energies where \begin{eqnarray}
z = E_{c.m.} + i \epsilon \equiv \frac{q_0^2}{m} + i \epsilon   .
\end{eqnarray}
We solved (\ref{2dimLSE}) by generating the corresponding Neumann series and
summing it up by Pade which is a very reliable and accurate method.
Usually six iterations were fully sufficient.
In each iteration the Cauchy singularity was split into a principal-value
integral (treated by subtraction) and a $\delta$-function piece. We used 120
or 140 $q$-points and 150 or 190 $x$-points. The $q$-integral points are
chosen in the definite interval $(0, \bar{q} )$, where typically $ \bar{q}$=
50 fm$^{-1}$.
In order to obtain directly the on-shell t-matrix 
element $ T( q_0, q_0, x, E_{c.m.}
)$ we added $q=q_0$ to the set of $q$-points. To better control the behavior
of the transition matrix element for small scattering angles also $x=1$ was
added to the set of $x$-points.
A typical run required less than 9 minutes on 256 nodes (1024 processors) on
the IBM Blue Gene/P parallel computer at the J\"ulich Supercomputing Centre.

In Figs.~\ref{fig3}-\ref{fig5} we show with dash-dotted line the 
real and imaginary parts of
the transition amplitudes
 $A_C (\theta) \equiv -2 \pi^2 m \, T( q_0, q_0, \cos \theta ) $ 
 for  sharp cut off Coulomb potential 
pp scattering at $E_p^{lab}=13$~MeV  
 and a number of cut-off radii $R=10$ and 
$20$~fm (Fig.~\ref{fig3}), $R=40$ and $80$~fm (Fig.~\ref{fig4}),  and 
 $R=100$ and $120$~fm (Fig.~\ref{fig5}). With increasing cut-off
radius a development of strong oscillations in the scattering angle 
dependence for the real parts of the numerical solutions is clearly seen. 
These oscillations follow on average the real part of the pure Coulomb
amplitude given by (\ref{eq52}) and shown by the solid line. 
The imaginary parts of the numerical
solutions are totally off from the imaginary part of the pure Coulomb
amplitude and have even an opposite sign. 
 Now applying to the numerical solutions the asymptotic renormalization
 factor from (\ref{eq53}) dramatically improves the agreement (dotted
 lines in Figs.~\ref{fig3}-\ref{fig5}). Not only the oscillations in
 the real parts are practically removed and the pure Coulomb and
 renormalized amplitudes are practically overlapping but the
 renormalization brings also imaginary parts into agreement with
 the exception of very forward angles.  
 When one desists to use the asymptotic expansion for $\tilde f_R$ and
 instead calculates it exactly according to (\ref{eq33}), (\ref{eq34})
 and  (\ref{eq43}) than the ratio $\frac {\tilde f_R} {A_C (\theta)}$
 provides the exact renormalization factor. Performing exact
 renormalization of the numerical solutions provides very good
 agreement between imaginary parts of the numerical and pure Coulomb
 amplitudes also at the very forward angles (dashed line in 
 Figs.~\ref{fig3}-\ref{fig5}). 

We also checked how important are the  two additional terms in the
renormalization factor of (\ref{eq53}). To this aim we 
renormalized the numerical solutions with the standard form of
the renormalization factor, given by the first term in
(\ref{eq53}). In Fig.~\ref{fig6} solid (red) lines show the amplitude
renormalized in this way. It is clearly seen, that restricting to the
standard form of the renormalization factor it is not possible to
reach the physical amplitude. Standard renormalization reduces
slightly oscillations in the real part of the numerical solution and
changing the sign of the imaginary part invokes in it  large
oscillations. So after standard renormalization strong oscillations are present
both in the real and imaginary parts and fails totally.

\section{Summary}
\label{summary}
The renormalization method for a screened on-shell Coulomb t-matrix enjoys a
widespread use; see for instance \cite{Alt02,Deltuva05}.  As pointed out in
the introduction the underlying
mathematical considerations leave room for doubts. To shed light on that
issue we regarded potential
scattering on a sharply cut-off Coulomb potential directly in 3 dimensions,
avoiding obstacles in the
infinite sum of angular momenta. The idea was to use the Lippmann-Schwinger
equation which uniquely defines
the wave function including its boundary conditions. Inside the range of the
potential it is the standard
Coulomb wave function multiplied by an unknown normalisation factor. Using
that form also on the
left side of the Lippmann-Schwinger equation for radii smaller than the
cut-off radius determines that
normalisation factor uniquely. Based on that we succeeded analytically to
determine the
normalisation factor  and thus obtained in this manner the exact analytic
result for the wave function.
 This also allowed us to derive the analytical expression for the scattering
amplitude in the limit of
infinite cut-off radius. The connection to the standard Coulomb scattering
amplitude $ A_c( \theta)$
turned out, however, to be different from the standard form used widely in
the literature and is
given in (\ref{eq53}). Our form  consists of two terms, one of which is the
standard one, $ e^ { -
2i \eta ln 2pr} A_c( \theta)$. To that, however,  is added a new expression
which is singular at $ \theta
=0$ and $\theta = \pi$.
 These analytical results are fully backed up by accompanying numerical
investigations. Our renormalization factor brings in a very good
agreement between the strongly deviating and 
oscillating numerical solution of the
Lippmann-Schwinger equation with the sharp cut  off Coulomb potential 
and the exact Coulomb amplitude.  The
standard renormalization factor fails completely.

\section*{Acknowledgments}
This work was supported by the 2008-2011 Polish science funds as a
 research project No. N N202 077435. It was also partially supported by the
Helmholtz
Association through funds provided to the virtual institute ``Spin
and strong QCD''(VH-VI-231) and by
  the European Community-Research Infrastructure
Integrating Activity
``Study of Strongly Interacting Matter'' (acronym HadronPhysics2, Grant
Agreement n. 227431)
under the Seventh Framework Programme of EU.
 The numerical calculations were
performed on the IBM Regatta p690+ of the NIC in J\"ulich,
Germany.

\clearpage

\appendix

\section{S-wave potential scattering for a sharply cut-off Coulomb
potential}
\label{apen1}

The (reduced) wave function for s-wave scattering obeys the Lippmann-Schwinger equation
\begin{eqnarray}
\phi^ {(+)}(r) = \sin(pr) - \frac{m}{p} \int_0^ R dr' e^ { i p r_>} \sin( p r_<)
\frac{e^ 2}{r'} \phi^
{(+)}(r')\label{eqa1}
\end{eqnarray}
with $ r_{<(>)}$ the smaller ( greater)  of $ r,r'$. Inside the potential
range $ \phi^ {(+)}(r)$ has to
have the form
\begin{eqnarray}
\phi^ {(+)}(r) = A F_0 ( p r)\label{eqa2}
\end{eqnarray}
where $ F_0 ( p r)$ is proportional to the standard Coulomb wave function
\begin{eqnarray}
F_0 (pr) =  pr e^ { i pr} F ( 1 + i \eta, 2, - 2ipr) ~.
\label{eqa3}
\end{eqnarray}

Inserting (\ref{eqa2}) into (\ref{eqa1}) yields
\begin{eqnarray}
\phi^ {(+)}(r) & = &  \sin(pr) - 2 \eta  A (   e^ { i p r} \int_0^ r dr' \sin(pr')
 \frac{1}{r'} F_0
(pr')
 +   \sin(pr) \int_r^ R dr' e^ { i p r'} \frac{1}{r'} F_0 (pr'))\cr
& = &  \sin(pr) - \frac{2 \eta p A}{2i} ( e^ { i p r} \int_0^ R dr'e^ { 2i p
r'} F ( 1 + i \eta, 2, -
2ipr')\cr
& - &  e^ { - i p r}\int_r^ R dr'e^ { 2i p r'}F ( 1 + i \eta, 2, - 2ip
r')
 -   e^ { i p r}\int_0^ r dr' F ( 1 + i \eta, 2, - 2ip r')) ~.
\label{eqa4}
\end{eqnarray}

One faces two types of integrals, which can be solved using the following
properties of the confluent
hypergeometric function:
\begin{eqnarray}
F ( 1 + i \eta, 2, - 2ip r) = \frac{1}{ 2 p \eta} \frac{d}{dr} F (  i \eta,
1,  - 2ip r)\label{eqa5}
\end{eqnarray}
\begin{eqnarray}
F ( 1 + i \eta, 2, - 2ip r) = - \frac{e^ { - \pi \eta}}{ 2 \pi \eta}
\int_{\Gamma} dt e^ { - 2iprt}
 (\frac{t}{1-t})^ {i \eta}\label{eqa6}
\end{eqnarray}
with the path $ \Gamma$ given in section \ref{scatamplitude}, and 
\begin{eqnarray}
 F (  i \eta, 1,  - 2ip r) - F ( 1 + i \eta, 1, - 2ip r) = 2 i pr F ( 1 + i
\eta, 2, - 2ip r) ~.
\label{eqa7}
\end{eqnarray}

One obtains
\begin{eqnarray}
  \int_0^ r dr'F ( 1 + i \eta, 2, - 2ip r') = \frac{1}{ 2 p \eta}  ( F (  i
\eta, 1,  - 2ip r)
-1)
\label{eqa8}
\end{eqnarray}
\begin{eqnarray}
\int_0^ r dr'  e^ { 2i p r'} F ( 1 + i \eta, 2, - 2ipr') =  - \frac{1}{ 2
\eta p}( 1 - e^ { 2 i pr}
 F ( 1 + i \eta, 1, - 2ipr) ~.
\label{eqa9}
\end{eqnarray}
Therefore
\begin{eqnarray}
\phi^ {(+)}(r) & = & \sin(pr) - \frac{2 \eta p A}{2i} ( \frac{ 2i}{ 2 \eta p}
\sin(pr) e^ { 2 i pR}
 F ( 1 + i \eta, 1, - 2ipR)
 -   \frac{ 2ipr}{ 2 \eta p} e^ { i pr} F ( 1 + i \eta, 2, - 2ipr))\cr
& = &  \sin(pr) ( 1 - A e^ { 2 i pR}  F ( 1 + i \eta, 1, - 2ipR)) 
 +   A p r e^ { i pr} F ( 1 + i \eta, 2, - 2ipr)\cr
& = &  \phi^ {(+)}(r) + \sin(pr) ( 1 - A e^ { 2 i pR}  F ( 1 + i \eta, 1, -
2ipR)) ~.
\label{eqa10}
\end{eqnarray}
Consequently the LS equation (\ref{eqa1}) is identically fulfilled, as it
should and one obtains an
explicit
condition for the constant A:
\begin{eqnarray}
1 - A e^ { 2 i pR}  F ( 1 + i \eta, 1, - 2ipR) = 0\label{eqa11}
\end{eqnarray}
or
\begin{eqnarray}
 A = \frac{e^ { - 2 i pR}}{  F ( 1 + i \eta, 1, - 2ipR)} ~.
\label{eqa12}
\end{eqnarray}

Inserting this result into (\ref{eqa2}) the exact s-wave function  for a
sharply cut-off Coulomb  is
obtained
\begin{eqnarray}
\phi^ {(+)}(r) & = & \frac{e^ { - 2 i pR}}{  F ( 1 + i \eta, 1, - 2ipR)} pr
e^ { i pr} F ( 1 + i \eta, 2,
- 2ipr)\label{eqa13}
\end{eqnarray}
It obeys the  LS equation (\ref{eqa1}).

The asymptotic behavior $ r \to \infty$, which provides the scattering phase
shift $ \delta_R(p)$, is given
through the LS
equation and we read off from (\ref{eqa4})
\begin{eqnarray}
\phi^ {(+)}(r) \rightarrow \sin(pr) - e^ { i pr} A'
\label{eqa14}
\end{eqnarray}
with
\begin{eqnarray}
 A' = 2 \eta p A \int_0^ R dr' \sin(pr') e^ { i pr'} F( 1 + i \eta, 2, - 2 i
pr') ~.
\label{eqa15}
\end{eqnarray}
At the same time this yields
\begin{eqnarray}
 e^ { 2 i \delta_R(p)} = 1 - 2i A' ~.
\label{eqa16}
\end{eqnarray}
Using (\ref{eqa8}) and (\ref{eqa9}) again gives
\begin{eqnarray}
A' = \frac{A}{2i} ( e^ { 2i pR } F( 1 + i \eta, 1, - 2 i pR) - F(  i \eta,
1, - 2 i pR))
\label{eqa17}
\end{eqnarray}
and consequently
\begin{eqnarray}
 e^ { 2 i \delta_R(p)} = 1 - A( e^ {2 i pR } F( 1 + i \eta, 1, - 2 i pR) -
F(  i \eta, 1, - 2 i
pR)) ~.
\label{eqa18}
\end{eqnarray}

The interest lies now in the limit $ R \to \infty$. We use
(\ref{eqa12}) and
the asymptotic form
(\ref{abrom}) of $F$
 and  obtain
\begin{eqnarray}
e^ { 2 i \delta_R(p)} \rightarrow e^ { 2i \sigma_0 - 2i \eta \ln(2pr)}\label{eqa19}
\end{eqnarray}
or
\begin{eqnarray}
 \delta_R(p) \rightarrow  \sigma_0 -  \eta \ln(2pr) ~.
\label{eqa20}
\end{eqnarray}

Of course this result is well known and can be trivially obtained by
matching the interior Coulomb wave
function to the free one containing $\delta_R(p)$.

We performed this exercise to explicitly demonstrate that the LS equation
(\ref{eqa1}) is indeed
identically
fulfilled for arbitrary $r$  below the cut-off radius $ R$. In the 
3-dimensional case we succeeded
analytically  to do this only for the special value $r=0$, though 
it is valid for any $r < R$, and were forced to
verify the general case
numerically.

\begin{table}
\caption{\label{table1} The left and right sides of (\ref{eq15}) at
  $E_p^{lab}=13$~MeV ($\eta=0.0439$, $p=0.3959$~fm$^{-1}$) 
and different screening radii $R$.
}
  \begin{tabular}{l c c}
   \hline
   \hline
   $R$~[fm] & $\tilde A$ & 
 $ 1 -\tilde  A \eta T \int_0^1 dx {_1F_1} ( - i \eta, 1,
iTx)\int_0^{1-x} dy e^{ iTy}\frac{1}{ x+y}$   \\
   \hline
0.5 & (0.98301, -0.00166) & (0.98301, -0.00166) \\
1   & (0.96724, -0.00634) & (0.96724, -0.00634) \\
5   & (0.91933, -0.08246) &  (0.91933, -0.08246)  \\
10 & (0.92770, -0.10294) & (0.92770, -0.10294) \\
20 & (0.91961, -0.13491) & (0.91961, -0.13491) \\
50 & (0.91606, -0.17185) & (0.91606, -0.17185) \\
100 & (0.90960, -0.20061) & (0.90960, -0.20061) \\
500 & (0.89376, -0.26439) & (0.89377, -0.26439) \\
1000 & (0.88528, -0.29140) & (0.88528, -0.29140) \\
5000 & (0.86250, -0.35307) & (0.86252, -0.35307) \\
   \hline
  \end{tabular}
\end{table}

\begin{table}
\caption{\label{table2} The exact value of $\tilde{A}$ as 
in (\ref{eq24}) (left column), 
asymptotic form  given by (\ref{eq511}) (middle column) and their 
ratio (right column) 
at $E_p^{lab}=13$~MeV for different screening radii $R$.
}
\begin{tabular}{r c c c}
\hline
\hline
$R$~[fm] & $\tilde{A}$ & ${\tilde{A}}_{approx}$ &  
$\tilde{A}/{\tilde{A}}_{approx}$ \\ 
\hline
0.1 &  (0.99653,-0.00007)  &  (0.92852, 0.07999) &  (1.06534,-0.09185) \\
1.0   &  (0.96724,-0.00634)  &  (0.93186,-0.01402) &  (1.03784, 0.00881) \\
2.0   &  (0.94156,-0.02249)  &  (0.93100,-0.04233) &  (1.01035, 0.02178) \\
3.0   &  (0.92536,-0.04353)  &  (0.93010,-0.05888) &  (0.99389, 0.01611) \\
5.0   &  (0.91933,-0.08246)  &  (0.92855,-0.07970) &  (0.99040,-0.00380) \\
10.0  &  (0.92770,-0.10294)  &  (0.92570,-0.10788) &  (1.00152, 0.00551) \\
20.0  &  (0.91961,-0.13491)  &  (0.92199,-0.13596) &  (0.99731, 0.00074) \\
50.0  &  (0.91606,-0.17185)  &  (0.91578,-0.17289) &  (1.00009  0.00115) \\
100.0 &  (0.90960,-0.20061)  &  (0.91011,-0.20064) &  (0.99946,-0.00008) \\
\hline
\end{tabular}
\end{table}

\begin{figure}[hp]\centering
\epsfig{file=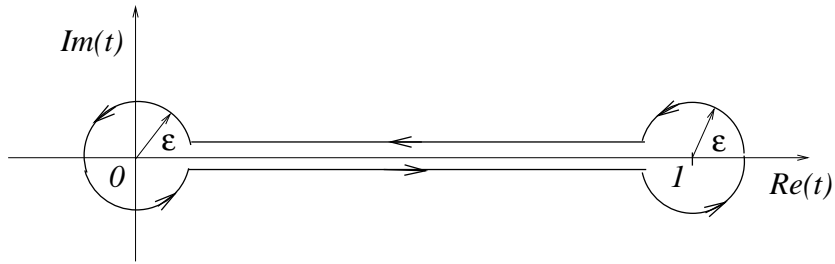,width=11cm}
\caption{The path of integration $\Gamma$ in Eq. (33)}
\label{fig1}
\end{figure}

\begin{figure}[hp]\centering
\epsfig{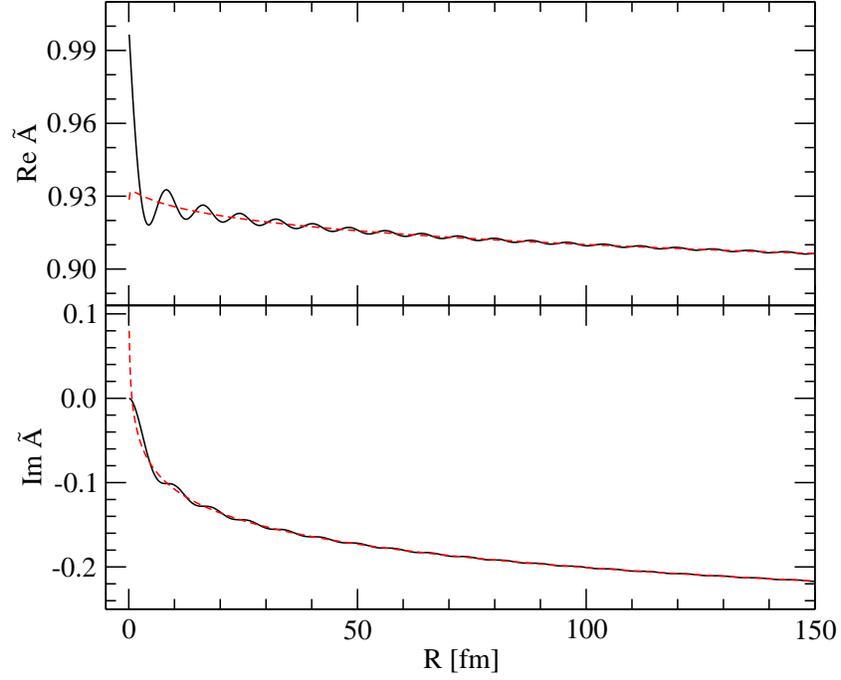}
\caption{(Color online) The real (top) and imaginary (bottom) part 
of $\tilde{A}$ as a function 
of the screening radius $R$ at $E_p^{lab}=13$~MeV. The solid (black) 
line represents 
the exact expression given in (\ref{eq24}) and the dashed (red) 
line shows the asymptotic 
form as given in (\ref{eq511}).}
\label{fig2}
\end{figure}

\begin{figure}[hp]\centering
\epsfig{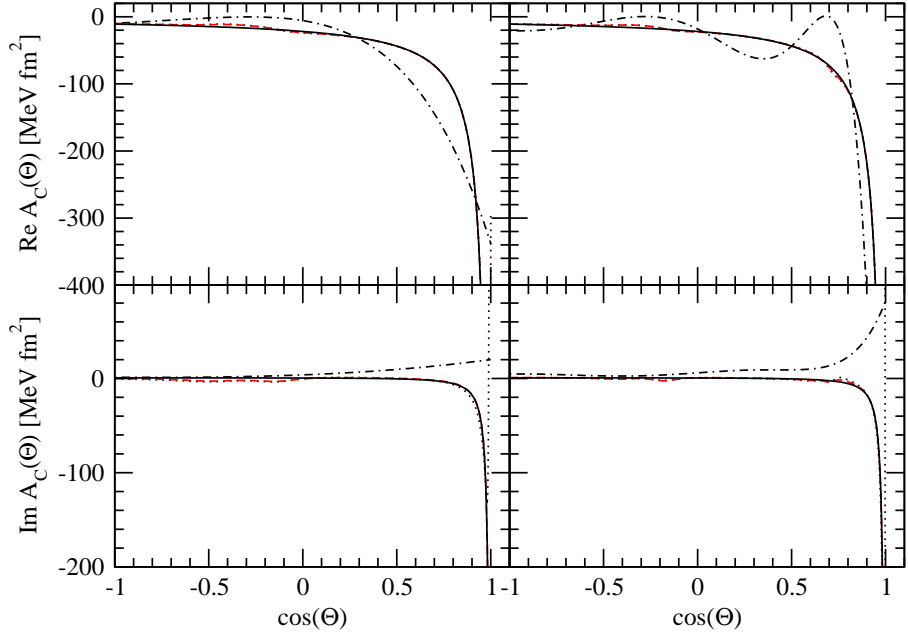}
\caption{(Color online) The real (top) and imaginary (bottom) part 
of $A_C (\theta) \equiv
-2 \pi^2 m \, T( q_0, q_0, \cos \theta ) $ as a function 
of $ \cos \theta $ for $R$= 10 fm (left panel) and 20 fm (right panel) 
at $E_p^{lab}=13$~MeV. 
The dash-dotted line represents a direct numerical prediction (without 
any renormalization).
The dotted line shows $A_C (\theta) $ with inclusion of the asymptotic 
renormalization factor 
given in (\ref{eq53}) and the dashed (red) line is for $A_C (\theta) $ with 
inclusion of the exact 
renormalization factor obtained from (\ref{eq33}), (\ref{eq34})
 and  (\ref{eq43}) (see text). The solid line represents the 
pure Coulomb amplitude given 
in (\ref{eq52}). Note that the dashed, dotted and solid lines 
practically overlapp with exception of very forward angles for 
imaginary part.} 
\label{fig3}
\end{figure}

\begin{figure}[hp]\centering
\epsfig{file=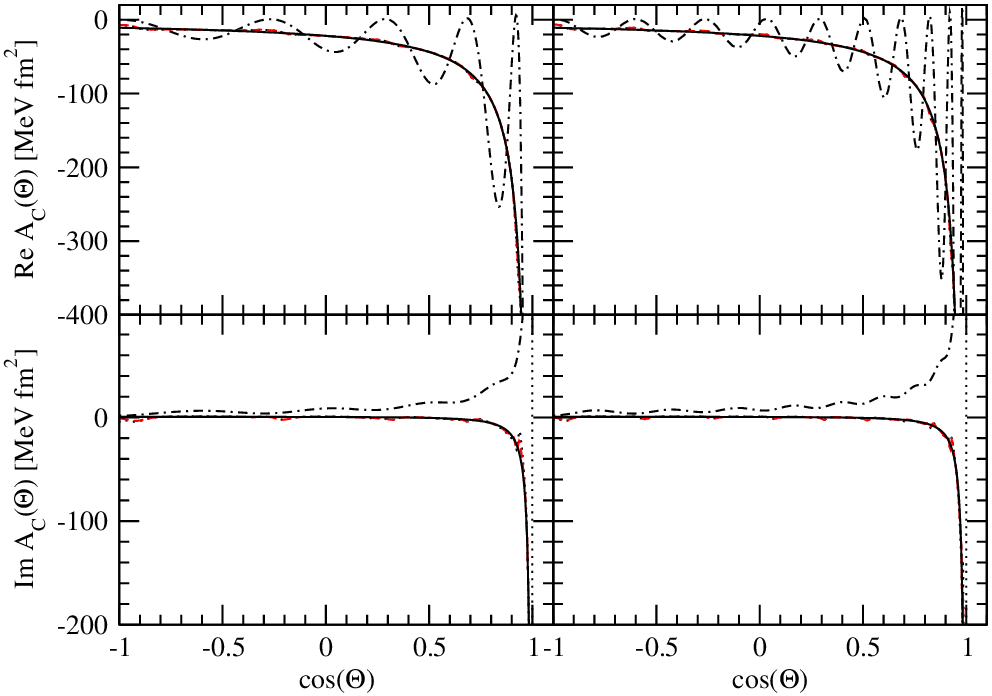,clip=true,width=12cm}
\caption{(Color online) The same as in Fig.~\ref{fig3} 
but for $R$= 40 fm (left panel) and 80 fm (right panel).}
\label{fig4}
\end{figure}

\begin{figure}[hp]\centering
\epsfig{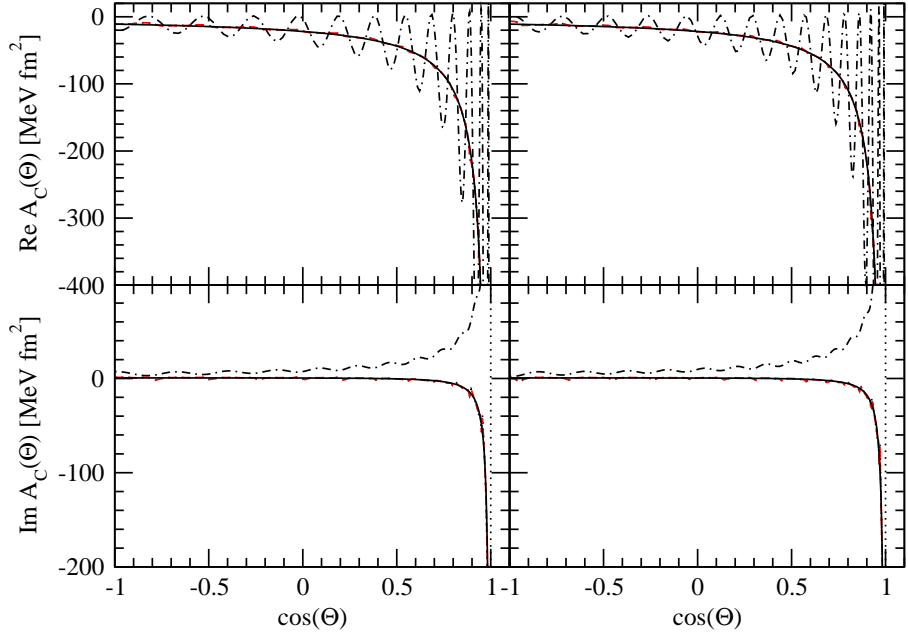}
\caption{(Color online) The same as in Fig.~\ref{fig3} 
but for $R$= 100 fm (left panel) and 120 fm (right panel).}
\label{fig5}
\end{figure}

\begin{figure}[hp]\centering
\epsfig{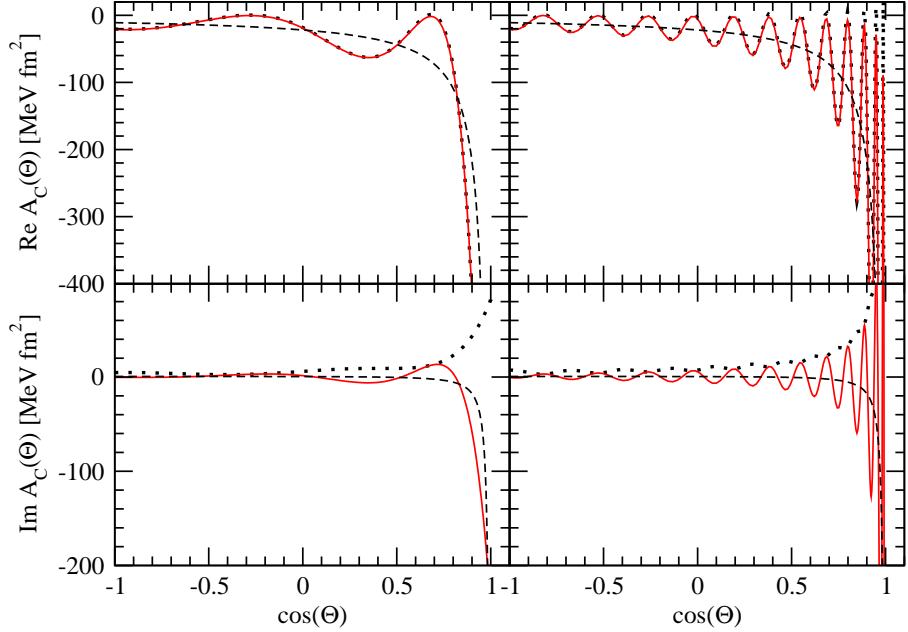}
\caption{(Color online) The real (top) and imaginary (bottom) part
of $A_C (\theta) \equiv
-2 \pi^2 m \, T( q_0, q_0, \cos \theta ) $ as a function
of $ \cos \theta $ for $R$= 20 fm (left panel) and 100 fm (right panel)
at $E_p^{lab}=13$~MeV.
The dotted line represents a direct numerical prediction (without
any renormalization).
The solid (red) line shows $A_C (\theta) $ with renormalization factor
$ e^ { - 2i \eta ln (2 p R)}$ and
the dashed line represents the
pure Coulomb amplitude given
in (\ref{eq52}).}
\label{fig6}
\end{figure}

\end{document}